# Gender Differences in International Research Collaboration in European Union

## Elsa Fontainha[1,2] and Tanya Araújo[1,2]

[1]*ISEG, Universidade de Lisboa, Lisbon, Portugal*
[2]*ISEG RESEARCH in Economics and Management, Portugal*

## Abstract

This paper investigates International Research Collaboration (IRC) among European Union (EU) countries from 2011 to 2022, with emphasis on gender-based authorship patterns. Drawing from the Web of Science Social Science Citation Index (WoS-SSCI) database, a large dataset of IRC articles was constructed, annotated with categories of authorship based on gender, author affiliation, and COVID-19 subject as topic. Using network science, the study maps collaboration structures and reveals gendered differences in co-authorship networks. Results highlight a substantial rise in IRC over the decade, particularly with the USA and China as key non-EU partners. Articles with at least one female author were consistently less frequent than those with at least one male author. Notably, female-exclusive collaborations showed distinctive network topologies, with more centralized (star-like) patterns and shorter tree diameters. The COVID-19 pandemic further reshaped collaboration dynamics, temporarily reducing the gender gap in IRC but also revealing vulnerabilities in female-dominated research networks. These findings underscore both progress and persistent disparities in the gender dynamics of EU participation in IRC.

---

⌂ Tanya Araújo
  tanya@iseg.ulisboa.pt
  R. Miguel Lupi 20, 1249-078 Lisboa, Portugal

# 1 Introduction

The growing International Research Collaboration (IRC) between individuals, teams, institutions and firms, countries and regions, is a key feature to solve complex scientific and societal problems like pandemics and climate change, to improve research excellence and efficiency and to accelerate knowledge innovation. Scientists from different countries cooperate in research activities or projects resulting in scientific outputs like patents or publications. Since the early research on IRC (Frame & Carpenter, 1979, Katz & Martin, 1997, Narin, et al., 1991), its drivers, barriers, effects, measurement methods, trends, patterns, and networks have been extensively analyzed.

International research collaboration has steadily increased globally across all countries and scientific fields over the past four decades, though its pace and intensity vary (Asknes, et al., 2019, Kwiek, 2021, Narin, et al., 1991). European framework programs have demonstrably boosted collaboration both within and outside Europe (Kwiek, 2021, Narin, et al., 1991). Historical trends reveal that linguistic, cultural, and historical factors significantly influence collaboration patterns, while emerging economies are increasingly contributing to global IRC (Adams, 2013, Coccia & Wang, 2016, Kwiek, 2021, Narin, et al., 1991). For the European Union (EU), the United States is the leading non-European partner when it comes to collaborations (Kwiek, 2021, Narin, et al., 1991). Although still lower than other fields, collaboration rates in Humanities and Social Sciences have shown significant growth (Asknes, et al., 2019, Kwiek, 2021, Narin, et al., 1991).

Despite the increasing importance of IRC for knowledge production, persistent gender disparities continue to exist, women still face disadvantages in terms of publication rates, representation in senior positions, citation counts, and recognition within academia (Fox, et al., 2017, Sugimoto & Lariviere, 2023, Uhly, et al., 2017). These disparities stem from family responsibilities acting as a *glass fence* and broader institutional and structural barriers (Uhly, et al., 2017, Zippel, 2019), with policy implementation varying significantly across EU (Anagnostou, 2022). Women's publications are often more domestically oriented, leading to fewer citations than men's, and their representation diminishes in higher academic ranks (Larivière, et al., 2013, Uhly, et al., 2017). The impact of inter-gender collaboration can have opposite effects: potentially benefiting male scholars while negatively affecting female scholars, particularly with high-level partners collaborators (Shen, et al., 2022 a, Shen, et al., 2022 b). The gender gap in IRC is more pronounced in the Sciences than Social Sciences, partly due to differing resource needs and team sizes (Asknes, et al., 2019, Kwiek & Roszka, 2021a, Kwiek & Roszka, 2021b, Kwiek & Roszka, 2022, Sugimoto & Lariviere, 2023). Gender homophily in collaboration and citation perpetuates the exclusion of female from male dominated networks (Kwiek & Roszka, 2022, Whittington, et al., 2024, Zippel, 2017, Zhou et al., 2024).

The COVID-19 pandemic reshaped IRC, emphasizing the need for rapid information sharing to combat the virus and its socio-economic impacts (Kwon, et al., 2023). While online meetings and computer-mediated experiments provided imperfect alternatives to disrupted traditional methods, they also highlighted gender disparities, with women facing amplified challenges due to increased family burdens and professional conflicts (Garc´ıa-Costa et al., 2024, Goldin, 2021, Goldin, 2022).

By utilizing bibliometric data from 2011 to 2022 collected from Web of Science (WoS), Social Science Citation Index (SSCI), and applying network techniques, this research aims to enhance the understanding of the IRC patterns among Social Science researchers in the EU, while also examining differences by gender. To achieve this goal, the study will address the following research questions:

1. How have patterns of international research collaboration, both intra-EU and with non-EU countries, developed in the context of EU scientific and innovation policies?

2. Are there any differences in IRC patterns based on gendered authorship categories?

3. Has the COVID-19 pandemic crisis altered the extent and nature of gender disparities in IRC, particularly in terms of intensity and collaborative patterns?

The remainder of the paper is structured as follows: Section 2 provides a brief literature review; Sections 3 and 4 detail the relevant database construction, sample composition, outlines the methodology and present the results. The final section concludes.

## 2 Related Work

Our focus in this section is to present a brief review of studies concerning four major aspects of IRC: underlying drivers, obstacles and effects; global patterns and trends; the role of gender differences within it; and the impact of the COVID-19 pandemic.

### 2.1 International Research Collaboration: Drivers, Obstacles and Effects

The current phase of scientific progress is characterized by international collaborations, and, unlike the previous three phases—individual, institutional, and national—this one emphasizes global partnerships, which drive advancements in science but also make it harder for countries to keep their scientific assets, intellectual property, and skilled researchers (Adams, 2013). Jonathan Adams coined the *fourth age of research* term, emphasizing the growing importance of cross-border cooperation for high-impact research.

The drivers of IRC can be broadly categorized into three main groups (Abramo, et al., 2013, de Frutos-Belizón et al., 2024): (*i*) Macro-level factors, such as country characteristics, reward systems, research policies, scientific fields, and geographical distance; (*ii*) Organizational or collective factors, including reputation and resource availability; and (*iii*) Individual factors, such as motivation, career stage, and gender. The specific influence of researcher gender will be explored further in subsections 2.3 and 2.4.

Research on the individual drivers of IRC (Wagner & Leydesdorff, 2005) has shown that these collaborations can largely be explained by the principle of preferential attachment, they might be driven more by the self-interests of individual scientists than by broader institutional structures or policies. This conclusion, suggest that individual motivations can outweigh macro-level and organizational or collective groups of drivers.

The macro-level drivers of IRC include national or supranational policies and strategies and, consequently, the dynamics and characteristics of IRC in EU are significantly shaped by strategies and policies at supranational levels. The European Research Area (ERA) Policy Agenda for 2022-2024 prioritized fostering a fair and reciprocal environment for international cooperation (European Commission, 2021a). It also focused on developing attractive and sustainable research careers, encouraging balanced talent flow, and promoting mobility across different countries, disciplines, and sectors within the ERA. Furthermore, the Agenda emphasized gender equality and inclusion. Building on this foundation, the subsequent ERA Policy Agenda for 2025-2027 (European Commission, 2025a) promotes the unrestricted movement of researchers, knowledge, and data throughout the EU and enhancing inclusive and intersectional gender equality across the ERA is a central structural policy within the Agenda. The EU Scientific Advice Mechanism recommends to improve collaboration across EU countries and at the global level for knowledge sharing and coordination (European Commission DGRI, 2025a). The heightened prevalence of IRC in EU is a result of a concerted effort at the national and EU levels to foster such partnerships through policy and funding (e.g. EU research grants typically require applicants to form consortia with partners from at least three countries (Kwiek, 2021)).

Since 2012, within ERA, gender equality and gender mainstreaming in research have been a consistent focus. Several reports provide information for gender disparities in research and innovation (R&I) (European Commission, 2021b, European Commission DGRI, 2025b) specifically about gender differences in research publications and inventions, and success rates for grant applications. Some authors criticize the results of these EU policies and strategies, pointing out that despite 20 years of comprehensive policy development on gender equality in research, innovation, and higher education, implementation has been uneven. While North European countries have actively advanced these policies, their Southern and Eastern counterparts have lagged, creating growing disparities in policy adoption and progress across the continent(Anagnostou, 2022). In fact, the Gender Equality Index, which aims to measure progress of gender equality across the EU and its Member States, reveals relatively slow progress towards gender equality, with significant variations across countries and domains (EIGE, 2024). Note that while the Index encompasses six domains (work, money, knowledge, time, power, and health), its *knowledge* domain is restricted to education and training, excluding research activities.

The benefits of IRC that are frequently emphasized include: the synergistic sharing of financial, human, and material resources — such as equipment, laboratory facilities, and specimens — which leads to significant cost reductions and heightened operational efficiency; and the potential to enhance research quality and impact, develop institutional and individual capacities, and collectively address pressing global challenges (Aksnes & Sivertsen, 2023, Gu J. et al., 2024, Velez-Estevez, et al., 2022).

There are also impediments and disadvantages associated with IRC (Vabø & Schmidt, 2023).

These include the considerable costs required for effective coordination and management, disparities between national research systems, differing levels of professional experience, linguistic and cultural barriers, and the challenges posed by geographical distance (Chen, et al., 2019, Fitzgerald, et al., 2021, Jeong, et al. 2014, Matthews, et al, 2020). Based on a large survey to scientists from countries with different level of development a study revealed the existence of political, logistical, and cultural barriers to IRC. Scientists frequently pointed to a lack of international funding continuity (Lima-Toivanen et al., 2025), difficulties with material and data sharing, inconsistent academic standards and prejudice faced by scholars from emerging and developing countries (Matthews, et al, 2020).

## 2.2 Trends and Patterns of IRC

Most studies on IRC have focused on specific periods, fields of research, or regions. A recent overview (Aksnes & Sivertsen, 2023) covering the period from 1980 to 2021, analyzes over 50 million publications from WoS and maps the evolution of IRC, concluding that: international collaboration has grown steadily over four decades; international collaboration rates have increased across all fields and countries, but the intensity and pace of growth vary; EU Framework programs have boosted collaboration among European countries; IRC is most prevalent in natural sciences, particularly in disciplines requiring large research infrastructures like Astronomy or Evolutionary Biology; Humanities and Social Sciences have lower collaboration rates but have seen notable growth over four decades; and internationally co-authored publications attract more citations than national ones.

Another recent study analyzing IRC across all R&D fields in EU countries from 2009 to 2018 investigated several key aspects: the volume and growth of IRC and its contrast with the US and China; collaboration patterns; the divide between EU-15 (members before the 2004 enlargement) and EU-13 (older vs. newer members); and field-specific trends (Kwiek, 2021).The research concluded the following: in 2018, over half of publications in most EU countries involved IRC; the EU surpasses the United States and China in IRC volume and citations; the United States and China demonstrate strong internal collaboration networks compared with EU countries; the United States is the dominant collaboration partner for most EU countries; older EU members (EU-15) show higher levels of IRC and citation impact compared to newer ones (EU-13); for major European systems (e.g., the UK, France, Germany), publication output is almost entirely attributable to IRC; and fields with lower levels of IRC include Humanities and Social Sciences.

A foundational article in the scientometrics literature on international collaboration examined papers published between 1977 and 1986 (Narin, et al., 1991). Analyzing scientific fields both targeted and not targeted by the programs of the European Community (the predecessor to the EU), the study revealed the following key findings: There was a steady increase in collaboration, both within and outside the European Community, in the targeted fields compared to non-targeted areas. By 1986, more than one-fifth of papers were co-authored within the same country, and over 10% involved non-Community collaboration. The study also found that patterns of links between certain countries were influenced more by linguistic, historical, and cultural factors than by scientific output alone. Finally, Europe's main competitors and leading non-European collaborators were identified as the United States (which was also the top non-European collaborator), Japan, Canada, and Australia. Another earlier research (Adams, 2013) that analyzed 25 million articles from the Institute for Scientific Information (ISI) database (later the WoS database) from 1981 to 2011, and examined several countries including six European ones, found the following: research output in developed countries (e.g., the US, UK, Germany) was increasingly driven by international collaboration; China, India, Brazil, and South Korea had seen a 20-fold increase in research output since 1981, with 75% of their research remaining domestic in 2011. An article studying the period 1997 to 2012 concluded that: IRC had grown significantly across all research fields; there was a convergence between basic and applied sciences; eleven leading scientific countries (with the top six being the United States, the United Kingdom, Germany, France, Italy, and Canada) represented more than half of the total IRC; and the share of these leaders in IRC had decreased over time, reflecting the growing contributions of emerging economies (Coccia & Wang, 2016).

## 2.3   Gender Differences in International Research Collaboration

Major stakeholders in science and education policy, international organizations, and publishers such as the EC (European Commission DGRI, 2025b, European Commission, 2021b), and (Elsevier, 2017, Elsevier, 2020) UNESCO (UNESCO, 2022), NSF (NSB, 2024, NSF, 2004), consistently report on international collaboration and gender equality in research and education. Across the EU-27, from 2013 to 2022, the average proportion of women authors on publications resulting from IRC, by field of R&D, shows that women are underrepresented, although there has been a slight increase in their participation. Differences exist by R&D field: in Social Sciences, it increased from 34.4% (2013-2017) to 37.4% (2018-2022), and the maximum value attained is in Humanities and the Arts (41.3% in 2018-2022) (European Commission DGRI, 2025b, European Commission DGRI, 2025c). Research that compares gender collaboration patterns in Europe with those in other major global regions, such as North America, Asia, and Latin America, reveals varying dynamics driven by both local contexts and broader trends in IRC (European Commission, 2021b).

The discussion of the benefits of global science and global academy recognizes the diversity within IRC and their participants and identifies the existence of a gender gap that emerged as a crucial explanation for observed gender differences in research impact (Zippel, 2017). Despite the increasing participation of women in academia (UNESCO, 2025) and research (European Commission, 2021b) studies have found that the participation rate of women versus men in internationally coauthored publications is lower. The investigation of the evolution of participation in IRC by gender, allows for the identification of trends and patterns across research fields and geographies. For example, specific studies have focused on individual countries like China (Geng, et al., 2022, Zhang, et al., 2020), Italy (Abramo, et al., 2013, Abramo, et al., 2019), South Korea(Jeong, et al. 2014), Poland (Kwiek & Roszka, 2021a) and Spain (de Frutos-Belizón et al., 2023).

International collaboration in academia is on the rise and increasingly important for knowledge production and innovation, yet persistent gender disparities are evident (Fox, et al., 2017, Uhly, et al., 2017). Some studies suggest that women's participation in IRC has relatively decreased, despite an overall increase in IRC activity (Gabster, et al., 2020, Liu, et al., 2022, Pinho-Gomes, et al., 2020). Several factors contribute to these gender disparities. Family responsibilities often act as a *glass fence* for women, hindering their access to international knowledge networks and elite academic positions (Uhly, et al., 2017). Beyond individual circumstances, institutional and structural barriers also worsen gender inequalities in international collaboration (Zippel, 2019). Addressing these disparities requires policies to broaden participation and mitigate implicit biases in research and innovation (Anagnostou, 2022, Fox, et al., 2017). However, despite the European Union's comprehensive gender equality policies in research, implementation varies significantly. Southern and eastern European countries, for instance, often lag due to a lack of coherent discourse on structural barriers and implicit bias (Anagnostou, 2022).

These gender disparities in IRC have tangible effects on research output and recognition. Women's publication portfolios are often more domestically oriented, resulting in fewer citations compared to men, who benefit more from international collaborations (Larivière, et al., 2013). Women's representation diminishes at higher academic ranks, partly due to their less frequent engagement in IRC compared to men (Uhly, et al., 2017). Inter-gender international collaborations tend to be less continuous than intra-gender collaborations and have mixed effects: they can improve research performance for male scholars but negatively impact female scholars, especially when collaborating with high-level academic partners (Shen, et al., 2022 a, Shen, et al., 2022 b).

The majority of the cross-disciplinary literature on the gender differences on IRC converges on the existence of a larger gender gap in the case of the Sciences when compared to the Social Sciences (Asknes, et al., 2019, Kwiek & Roszka, 2021a, Kwiek & Roszka, 2022). Several causes may be at the origin of this gap: *(i)* In general, females have a larger share of publications in Social Sciences fields compared to Science fields. However, it is Science that predominates in IRC (Sugimoto & Lariviere, 2023); *(ii)* The financial and material resources necessary to carry out empirical analysis and experiences are lower in Social Sciences compared with Science and subsequently one of the motivations for IRC is weaker (Smykla & Zippel, 2010); *(iii)* The research teams are generally smaller in the case of Social Sciences and thus, also in bibliometric analysis of authorship, the average number of authors per article is lower and the cost of team coordination is smaller.

The diversity within research teams positively influences the quality and impact of research (European Commission DGRI, 2025b, Yang, et al., 2022). However, some researchers tend collaborate with (Kwiek & Roszka, 2021b, Kwiek & Roszka, 2022) or cite authors of the same gender (Zhou et al., 2024). Multiple factors influence research collaborations of the same gender (Kwiek & Roszka, 2021b, Zippel, 2017): resources, gender stereotypes, institutional environment, gendered specialization in research topics, reputation and perceived scientific performance. Homophily can arise either from the structure of the respective population (e.g.

disciplinary homophily in AI (Hajibabaei et al., 2022)) or from people's conscious or unconscious choices. This gender homophily is evaluated as a crucial mechanism that perpetuates the exclusion of women from networks where men are the majority (Kwiek & Roszka, 2021b). In EU-28, during the period 1989-2013, men and women exhibit homophilous attachment to authors of the same gender (Whittington, et al., 2024).

## 2.4 COVID-19 Impact on International Research Collaboration

The recent coronavirus pandemic has made it urgent and essential to share updated and accurate information and research results, to learn about the virus, to prevent and combat its spread, to find treatments and to identify the socio-economic negative impacts of pandemic crisis and mitigate them. All these factors incentivize IRC. At the same time, practical conditions for IRC have changed due to the confinement of universities and laboratories, as well as the drastic reduction in international travel, which has affected the mobility of researchers (e.g. participation in scientific meetings, internships, and scientific visiting programs), especially for women (Kwon, et al., 2023).

The COVID-19 pandemic suddenly imposes new forms of IRC as online meetings and computer mediated experiments. These new forms, which are not perfect substitutes for the previous ones, have opened up potential opportunities for those who previously had more difficulty with international mobility due to a lack of resources or professional and family commitments during the pandemic. However, extended work hours interfered with the academic pursuits of some professionals (Deryugina, et al., 2021, Gabster, et al., 2020, Pinho-Gomes, et al., 2020). The family burdens, in particular at households with dependent persons, were exacerbated with the closures of kindergarten, schools, home care services and other household support services (Garc´ıa-Costa et al., 2024).

Nobel laureate Claudia Goldin (2021, 2022) described the pandemic as a *she − cession* noting that women were more severely impacted than in previous recessions (Goldin, 2021, Goldin, 2022). This impact also extended to academic women and researchers. While there has been research on the impact of COVID-19 on gender inequalities in academic productivity (Lee, et al., 2023, Li et al., 2025), few studies have offered a global overview of the coronavirus pandemic's effects on gender disparities related to IRC in Europe.

A study investigating gender inequalities from several angles (authorship roles, leadership positions in publications, gender makeup of collaborative teams, and scientific impact), adopting difference-in-differences models and analyzing 33,104 biomedical and life sciences coronavirus-related papers published between 2018 and 2020 and 16,429 authors has relevant conclusions about collaboration (Liu, et al., 2022). The authors research 50 countries globally, without focusing on regional differences such as Europe or explicitly discussing IRC. They conclude that pandemic amplified existing gender inequalities in science, particularly in leadership, collaboration, and scientific impacts. There was an increase in publication counts, with a 25% increase for females and a 54% increase for males and simultaneously, a decrease the proportion of females' authorships relative to total authorships by 5% and consequently a gender gap widening.

There was a trend to gender homophily in co-authorships, especially among men: publications by only-male teams increased by 5% and only-female teams by 1.8%, whereas the share from mixed-gender teams declined by 6.8% (Liu, et al., 2022).

# 3  Data

The original bibliographic data was downloaded from Clarivate Analytics, Web of Science (WoS), Social Science Citation Index (SSCI). Only articles and review articles - hereafter referred to as 'articles' or 'papers'- were selected. The analysis focuses in particular on the period from 2011 to 2022. For each year, a random sample of 5,000 articles was selected, including all fields available in WoS database. Four types of information were added to each of the original articles collected from WoS: (i) whether the authors were affiliated with research institutions based in EU member countries and which countries they were from; (ii) for the years 2020 to 2022 if the article is about coronavirus or not; (iii) whether the article corresponded to an international collaboration; (iv) the gender of the authors.

The measure of IRC in this paper is based on co-authorship. This measure is frequently used as a proxy for measuring research collaboration, specifically IRC (Aksnes & Sivertsen, 2023, Wagner & Leydesdorff, 2005). Although acknowledged to have limitations (e.g. co-authorship only partially captures the extent and nature of the collaboration), co-authorship persists as the most useful and empirically robust indicator for assessing and quantifying patterns of research collaboration (Aksnes & Sivertsen, 2023, Katz & Martin, 1997).

The identification of the articles with participation from authors located in any of the EU countries was based on the next query:

QUERY01: **Articles with authors affiliated in a EU Country**

*Source Data Base*: Web of Science, Social Science Citation Index (SSCI).

*Query* : year, article, countries (current members of European Union + England, Wales, North Ireland and Scotland)

*Command for each year* : (PY=('year') AND DT=(article) AND CU*= countries in Tab.1 separated by the operator "OR" [1].

The identification of the country is based on authorship using the institutional addresses for each author available in the original database. If the author has more than one country address, it means, is affiliated in institutions from different countries the association is made to the country of the first affiliation institution (for a discussion of the multi-affiliation authorship issues see (Lin, et al., 2025) to understand the phenomenon of multi-affiliation in scientific authorship, its prevalence, implications, and how it varies by country and field during last decade). In the Web of Science database, a *Topic* query searches the Title, Abstract, Keywords, and Keywords Plus. To identify if the coronavirus pandemics is one of the topics of the article the Query02 was used.

---

[1]Note that the Field Tag "CU=Country/Region", searches for countries/regions in the Addresses field within each record of the WoS SSCI database.

Because the number of non-EU countries is very large and some have a very low scientific production in collaboration with researchers in EU a selection of countries to be studied was necessary. For the 178 countries whose authors have collaborated with EU, the average annual number of articles was computed. After, ordering the countries by the mean of articles per year (in decreasing order), the first 50 countries were selected from the total of 178. The highest yearly average belongs to the USA and the lowest to Slovenia. The selected 50 countries are listed in Table 1.

Table 1: List of Countries

| USA | Sweden | Scotland | Ireland | Saudi Arabia |
|---|---|---|---|---|
| England | South Korea | South Africa | Iran | Colombia |
| PR China | Brazil | Israel | Chile | Romania |
| Australia | Switzerland | India | Mexico | Hungary |
| Germany | Japan | Finland | Malaysia | Argentina |
| Canada | Belgium | New Zealand | Wales | Thailand |
| Netherlands | Taiwan | Poland | Greece | Nigeria |
| Spain | Norway | Portugal | Russia | North Ireland |
| Italy | Turkey | Austria | Czech Republic | Indonesia |
| France | Denmark | Singapore | Pakistan | Slovenia |

Web of Science Clarivate database includes, for recent years, the given name of the authors, which enables partially the gender identification of the authors. The gender disambiguation of the authors was made using a dataset available in Stanford Network Analysis Platform (SNAP) combined with the repository file nam-dict.txt, following the methodology adopted in (Araújo & Fontainha, 2018). Because the full name of the authors was missing for the years between 1999 and 2010, the study of gender authorship is restricted to 2011 to 2022.

## 3.1   Overview of the data set

A database was compiled comprising all articles that had at least two authors from different countries, with the restriction that at least one author must be from the EU (Query01). In order to isolate a subgroup of articles related to COVID-19 topic from 2020 to 2022, the Query02 was executed.

Articles in the Social Science Citation Index (SSCI) from the Web of Science (WoS) database with at least one author from the EU and co-authored by at least two countries show a large increase in the last years (Fig.1). Fig.2 shows the number of articles for the top 8 countries in number of articles, from 2011 to 2022. England stands out regarding EU's IRC, however, the maintenance of England in the same relative position is accompanied by a decrease in publications in 2020 and 2022 (Fig.2).

The trend in England publications appears to shift after 2019, a change that can likely be attributed to the decrease in its participation in EU projects following the UK's withdrawal from the EU (Bauzá et al., 2020, Highman, et al.,2023). Brexit, which officially began in January 2020 with a transition period until the end of that year, has significantly impacted global IRC for the UK, and this impact is primarily due to the weakening of its ties with EU research networks and funding programs, specifically Horizon 2020 (2014-2020) and Horizon Europe (2021-2027). Consequently, the UK's historical role as a central hub in EU research has declined after 2019, leading to reduced network centrality (Highman, et al.,2023).

Among the current EU countries, Germany and Spain have the second and third position, although the latter has been decreasing its participation in IRC in recent years. The United States, as expected, stands out as the most prominent among the non-EU countries, despite a decrease in absolute values since 2016 (Fig.2).

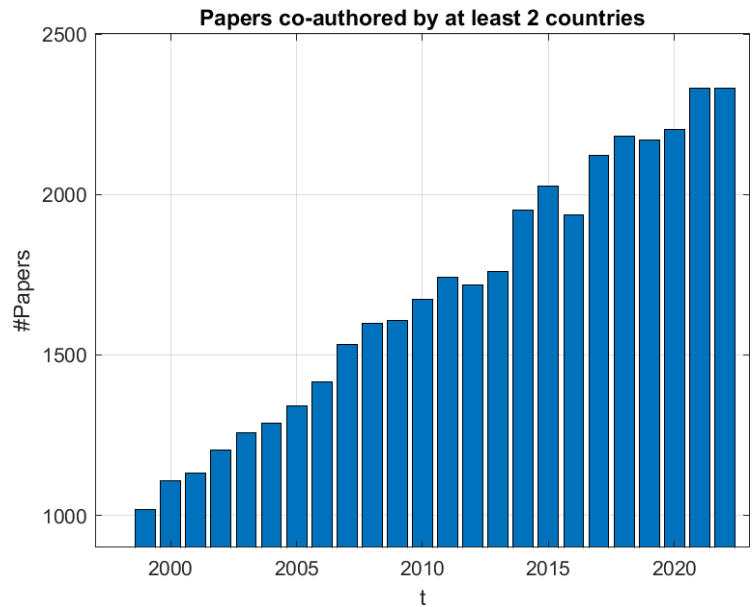

Figure 1: Articles with at least one author from EU and co-authored by at least two countries (1999-2022)

## 3.2 Authorship Categories

The definition of the five categories of authorship based on gender settles the basis for the identification of patterns of research collaboration and their relation to gender. Depending on the authorship characteristics, each paper belongs to at least one of the following categories. The categories are not mutually exclusive. In fact, the sum of categories 2 and 4 and the sum of categories 3 and 5 are equal, corresponding to Total, the category 1. Fig.3 shows the evolution of the number of authors in each category.

1. Total: all articles with at least two authors

2. Female: articles with at least one female author

3. OnlyFemale: articles exclusively authored by female

4. OnlyMale: articles exclusively authored by male

5. Male: articles with at least one male author

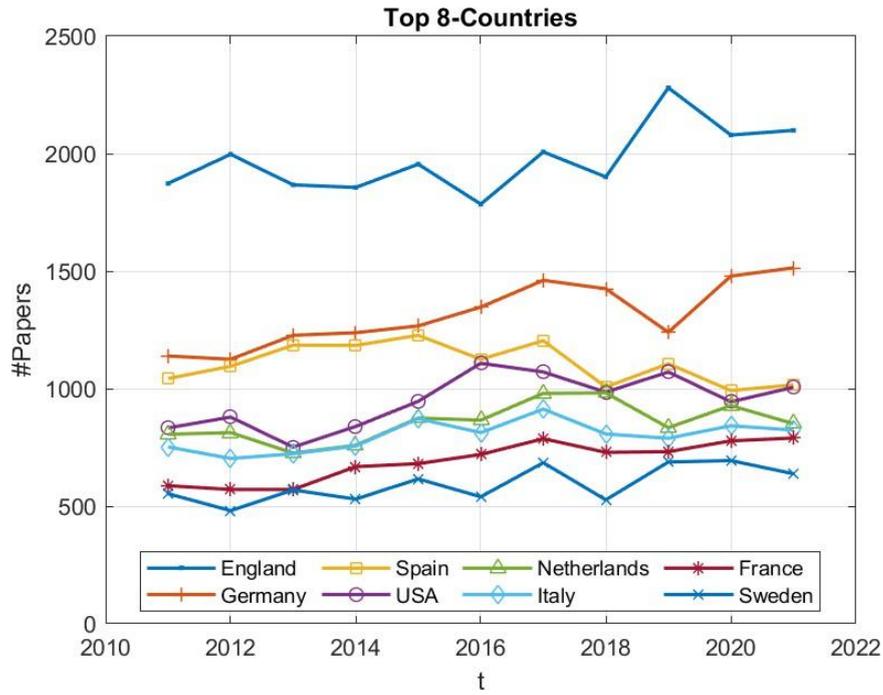

Figure 2: The top-8 countries in number of articles (2011-2022)

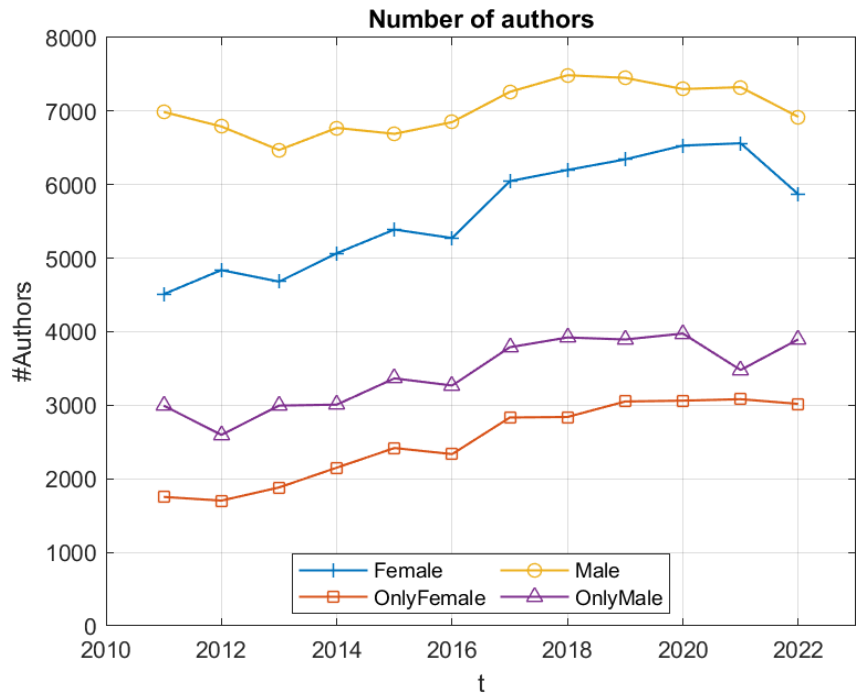

Figure 3: Authors by authorship category (2011 and 2022)

# 4  Methodology

Network induction refers to the method by which networks are defined from a certain data set. Network approaches are quite common in the analysis of systems where a network representation is the most intuitive one. And because connecting the elementary units of a system may occur in many different ways, that choice depends strongly on the available empirical data and on the questions that a network analysis aims to address (Araújo & Banisch, 2016).

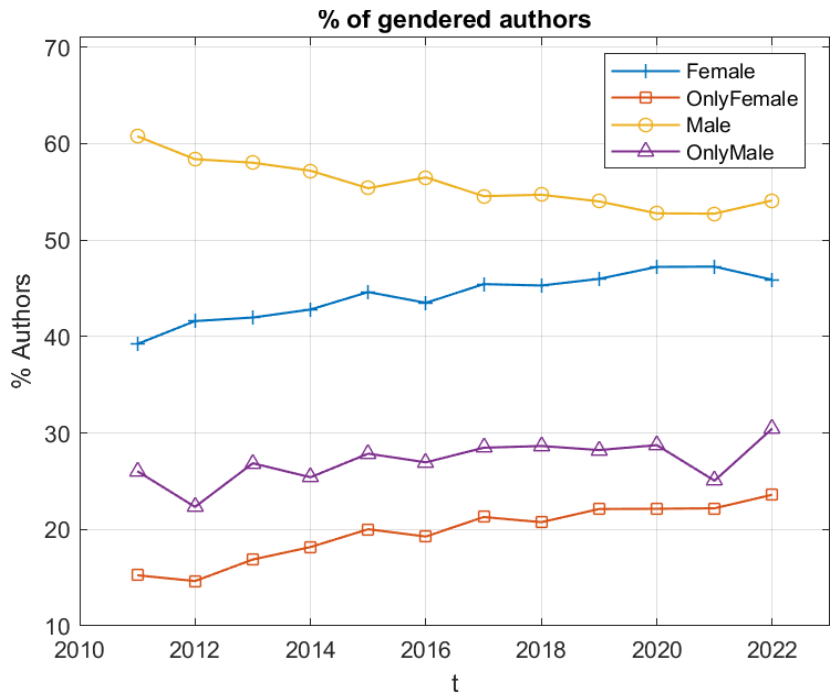

Figure 4: Percentage of Authors by authorship categories (2011-2022)

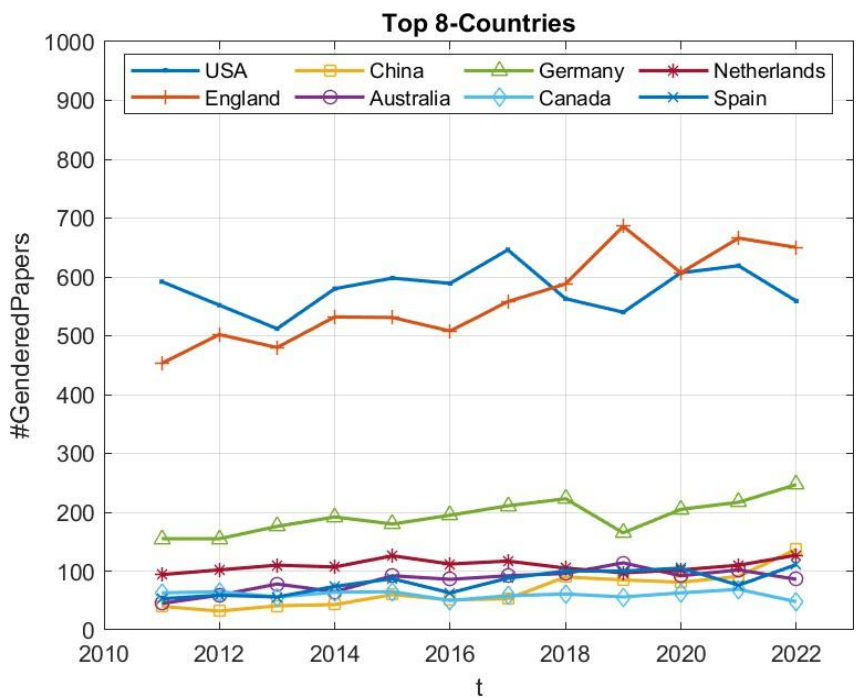

Figure 5: The top-8 countries in number of gendered articles (2011-2022)

Here, bipartite networks are induced from the subsets of articles defined by their co-authorship by countries. The frequency of co-occurrence of each pair of countries defines the existence of every link in the networks by authorship. The resultant networks are therefore weighted graphs since the weight of each link corresponds to the frequency of co-occurrence of the linked pair of countries.

In the next section, the complete networks are further analyzed through the construction of their corresponding minimal spanning trees (MST). In so doing, we are able to emphasize the main topological patterns that emerge from each network representation and to discuss their interpretation and relation to gender.

## 4.1 Connecting countries

A bipartite network $N$ consists of two partitions of nodes $V$ and $W$, such that the edges connect nodes from different partitions, but never those in the same partition. A one-mode projection of such a bipartite network onto $V$ is a network consisting of the nodes in $V$; two nodes $v$ and $v'$ are connected in the one-mode projection, if and only if there exist a node $w \in W$ such that $(v, w)$ and $(v', w)$ are edges in the corresponding bipartite network ($N$). In the following, we explore bipartite networks and their corresponding one-mode projections.

Each bipartite network by authorship category consists of the following partitions:

- the set $S$ of 50 countries presented in and

- one set of articles ($P_t^k$) by authorship category ($k = \{1, 2, 3, 4, 5\}$) as in Section 2.2 and $t = \{2011, 2012, ..., 2022\}$.

In each network ($N_t^k$), two countries are linked if and only if they co-occur in at least one article of $P_t^k$. Notice that each article $p \in P_t^k$ has at most ten authors. Besides, in each $p$, there shall be at least two authors affiliated to two different countries, one of which, from EU. A last attribute of each $p \in P_t^k$ concerns the identification of the gender of its authors. Naturally, the links in each network ($N_t^k$) are weighted by the number of coincident papers a pair of countries share in $P_t^k$.

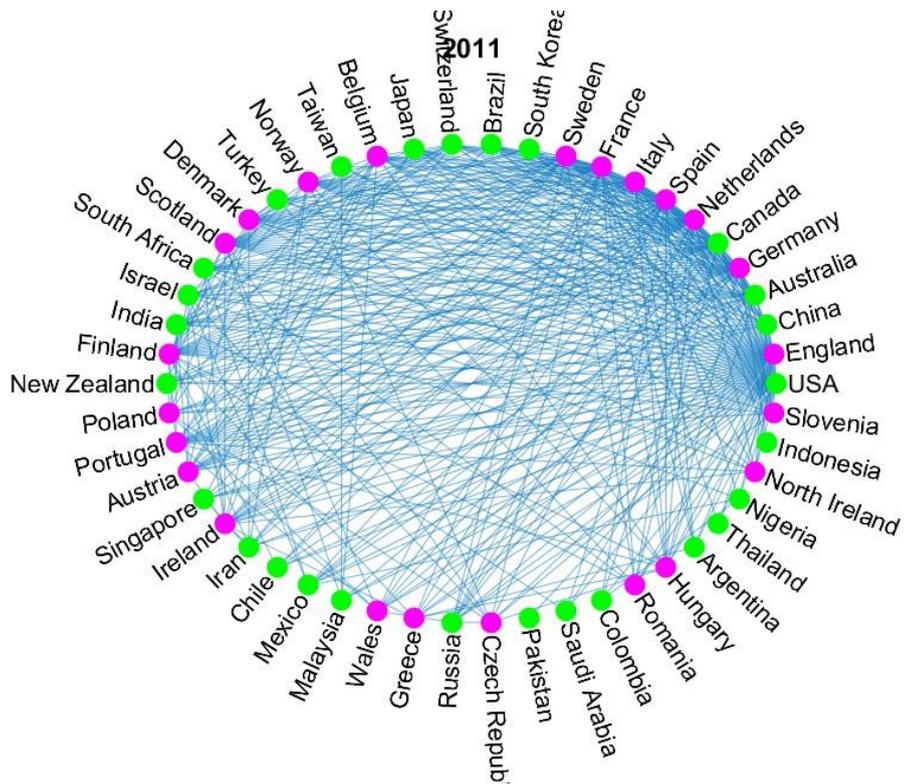

Figure 6: Complete network for 2011

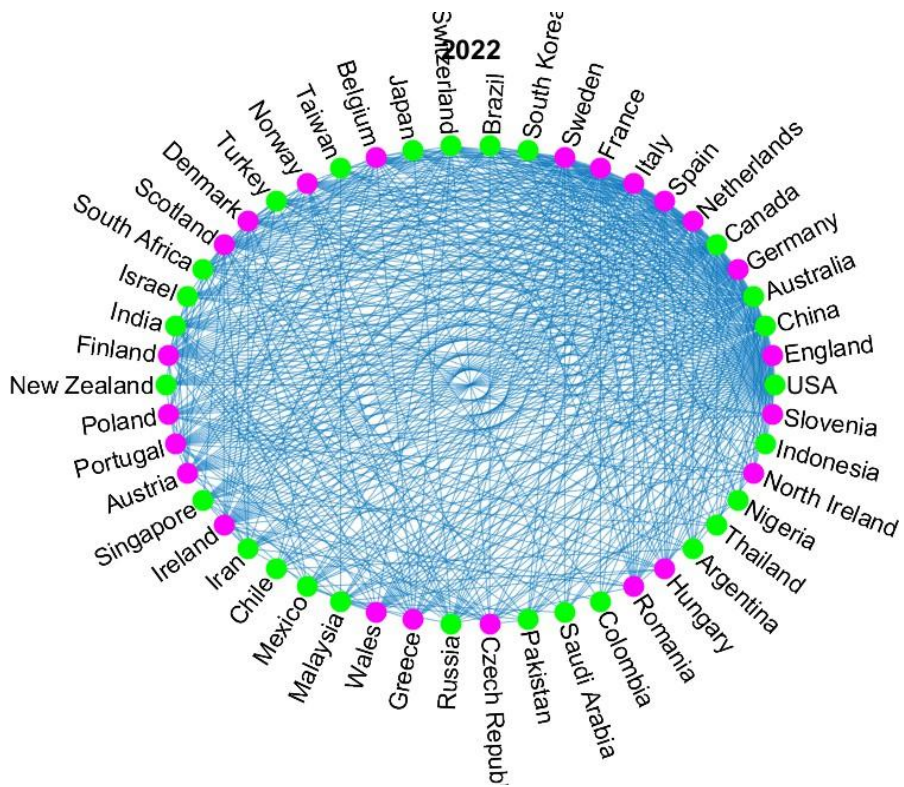

Figure 7: Complete network for 2022

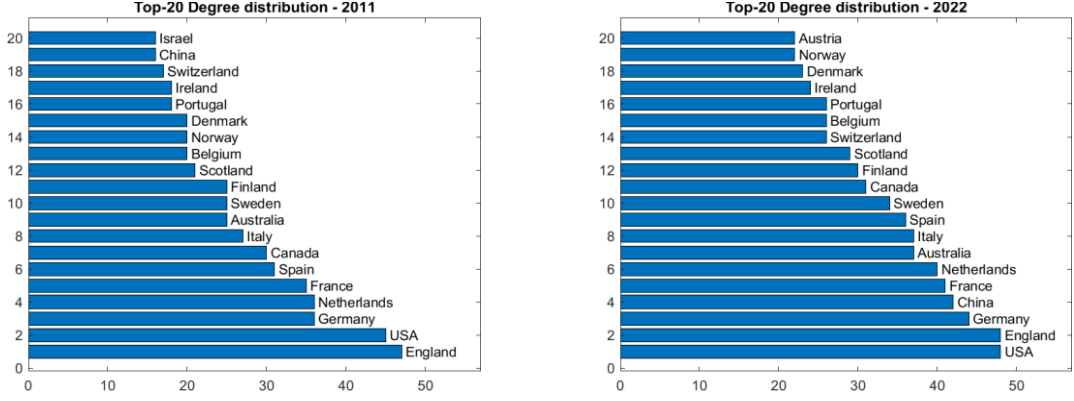

Figure 8: Degree distribution of the complete networks for 2011 and 2022

Fig.6 and Fig.7 show the complete and weighted networks induced from $P_{2011}^{Tot}$ and $P_{2022}^{Tot}$, i.e. induced from 2011 and 2022, respectively, set of papers. There, different colors are assigned to sets of nodes sharing the same community in the network, *pink* to EU countries and *green* for the non-EU countries. The networks in Fig.6 and Fig.7 display important differences between 2011 and 2022, being the most relevant ones:

1. the 2022 network displays a significantly more homogeneous degree distribution, indicating a more uniformly intensified collaboration across the network (see Fig.8).

2. in 2022, China emerged prominently in collaboration with England, which was accompanied by an intensification of scientific production by China.

The dominant positions of the US and England align with the findings of a previous study (Adams, 2013) where Adams's showed a significant increase in the percentage of international papers among total papers published annually: from 6% to 33% for the US, and from 10% to 52% for the UK. And China also saw an increase, moving from 10% to 25%.

In the next section, those weighted networks are further analyzed through the construction of their corresponding Minimal Spanning Trees (MST). In so doing, we are able to emphasize the main topological patterns that emerge from each network representation and to discuss their interpretation.

## 4.2 Minimal Spanning Trees

In the construction of a MST by the *nearest neighbor method*, the 50 countries are defined as the nodes ($n_i^k$) of a weighted and connected network ($N^k$) where the distance $d_{ij}^k$ between each pair of countries $i$ and $j$ corresponds to the inverse of weight of the link ($d_{ij}^k = \frac{1}{L_{ij}^k}$) between $i$ and $j$.

From the $n \times n$ distance matrix $D_{i,j}^k$ a hierarchical clustering is performed using the *nearest neighbor method*. Initially $n$ clusters corresponding to the $n$ countries are considered. Then, at each step, two clusters $c_i$ and $c_j$ are clumped into a single cluster if

$$d^k\{c_i, c_j\} = \min\{d^k\{c_i, c_j\}\}$$

with the distance between clusters being defined by

$$d^k\{c_i, c_j\} = \min\{d^k_{pq}\} \text{ with } p \in c_i \text{ and } q \in c_j$$

This process is continued until there is a single cluster. This clustering process is also known as the *single link method*, by which the MST of a graph is obtained. In a connected graph, the MST is a tree of $n-1$ edges that minimizes the sum of the edge distances. In a network with $n$ nodes, the hierarchical clustering process takes $n-1$ steps to be completed, and uses, at each step, a particular distance $d^k_k \in D^k$ to clump two clusters into a single one.

Let $C = \{d_q\}$, $q = 1, ..., N-1$, be the set of distances $d^k_{i,j} \in D^k$ used at each step of the clustering, and $thr = \max\{d_q\}$. It follows that $thr = d^k_{N-1}$.

Since some network-specific characteristics could come out from gendered authorship, here we consider the subsets of papers defined by the authorship categories presented in Section 2.2.

Minimum spanning trees allow for the identification of at least four important aspects that are not directly stated in complete networks (Araújo & Fontainha, 2017).

1. Branches: the way nodes organize themselves in different ramifications of the tree

2. Motifs: the prevalence of *star* motifs and/or *path* motifs in the tree

3. Connectivity: highly connected and weakly connected nodes

4. Centrality: the nodes occupying highly central positions and, conversely, those occupying the leafs of the tree

### 4.2.1   Tree motifs

The adoption of a network approach provides well-known notions of graph theory to fully characterize the structure of the networks. Here, and since our analysis relies on the MST, we concentrate on the calculation of just two topological coefficients, both measured at the network level.

The first one is the number of leafs ($l$) in the MST, i.e., the number of nodes with degree one. The second coefficient is the MST diameter ($d$), measuring the shortest distance between the two most distant nodes on the tree. The choice of these coefficients allows to characterize tree motifs with different shapes: from a pure *star* to a pure *path* motif.

It so happens that when the number of nodes of the tree is greater than 2, and depending on the motif that the MST approaches, its diameter ranges in between 2 and $N-1$ ($2 \leq d \leq N-1$). The closer is $\frac{d}{N-1}$ to 1, the smaller is the similarity of the MST to a *star* motif. Moreover, the number of leafs ranges in between exactly the same values but in the opposite direction, the closer $l$ is to 1, the smaller is the similarity of the MST to a *path* motif.

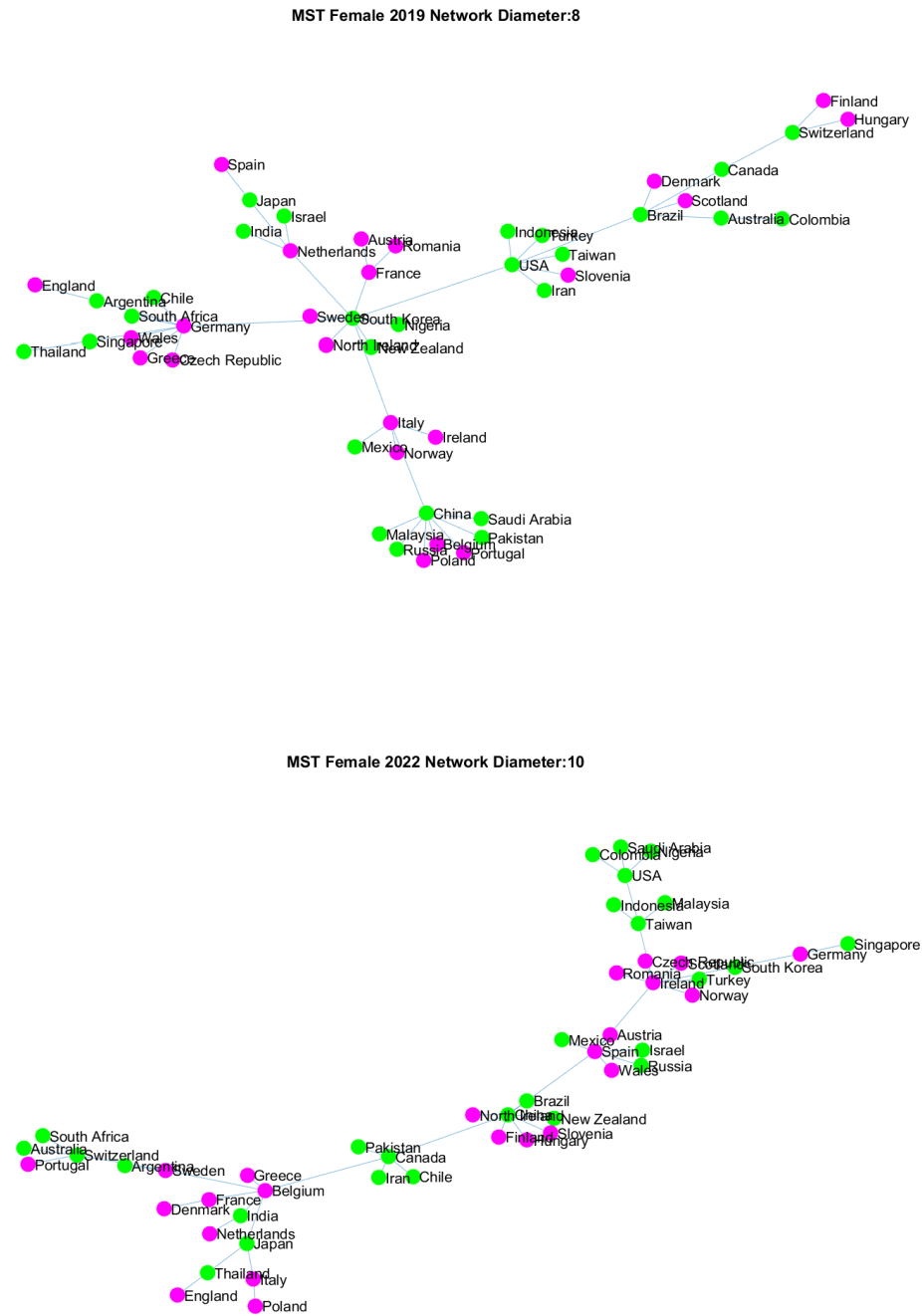

Figure 9: MST for the Female authorship category in 2019 and 2022

## 4.3 Results and Discussion

The trees induced from $P^{Female}_{2019}$ and $P^{Female}_{2022}$, i.e., induced from the set of articles where authors include at least one woman are shown in Fig.9. Likewise, Figure 10 shows the trees induced from $P^{Male}_{2019}$ and $P^{Male}_{2022}$ i.e., induced from articles where authors include at least one man.

The main differences in the two trees for years 2019 and 2022 presented in Fig.9 rely on:

- countries like South Korea, USA and Germany have their degree centrality decreased

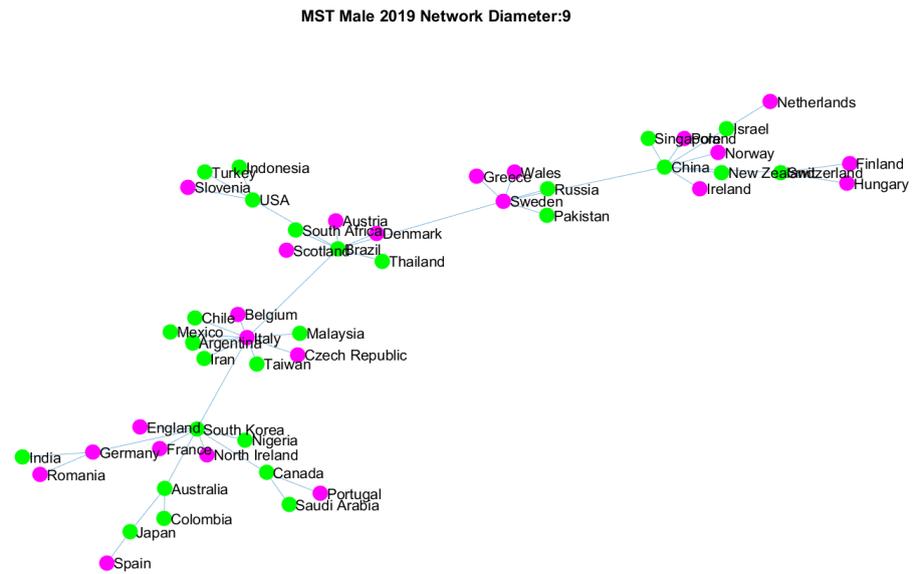

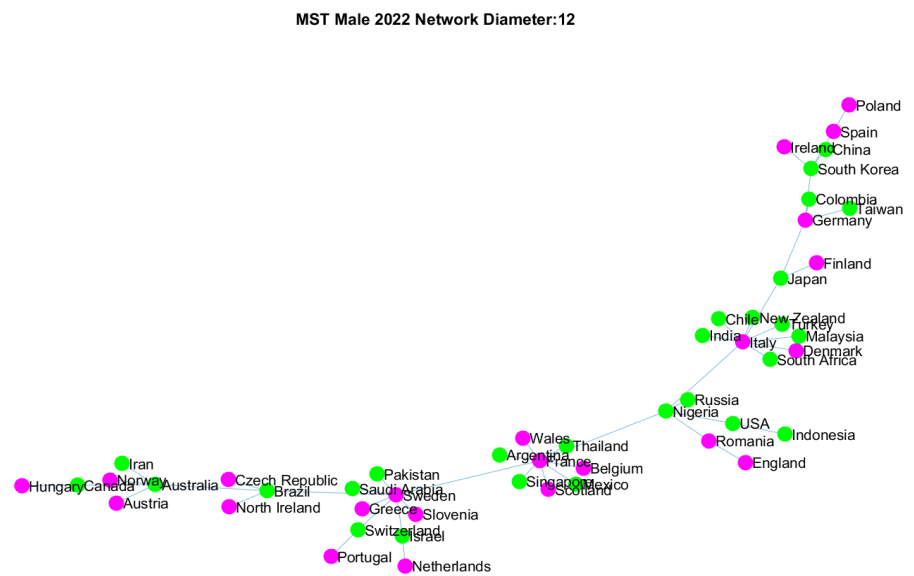

Figure 10: MST for the Male authorship category in 2019 and 2022

The female inclusive MST in 2022 has an increase in its diameter, as the values in Tab.2 show. Looking at the value of 2021, there is an increase of 50% in the Female MST's diameter when compared with the two previous years. Such a large increase is twofold: it highlights the role of China as an emerging hub in 2019. Such a role is weakened in 2022, leading the MST to a larger diameter and a more homogeneous degree distribution, as the bottom plot in Fig.9 shows.

| Authorship Category | 2019 | 2020 | 2021 | 2022 |
|---|---|---|---|---|
| **Total** | 7 | 10 | 13 | 8 |
| **Female** | 8 | 8 | 12 | 10 |
| **Male** | 9 | 10 | 8 | 12 |
| **OnlyFemale** | 8 | 8 | 7 | 7 |
| **OnlyMale** | 10 | 9 | 10 | 8 |

Table 2: MST diameter by Category and Year

Table 2 shows the values of the diameter - the shortest distance between the two most distant nodes - of each MST authorship category. Looking at these values, one sees that the greatest difference concerns 2021 values of the Total category when compared with OnlyFemale category. The latter displays a much small diameter and this is due to the fact that when male authorship is excluded, the topological distances among countries on the tree are reduced and the tree structure approaches a star-like motif. Such a structural difference leads to a less homogeneous distribution of the degrees of the nodes of the OnlyFemale MST.

Since the networks have the same size (50 countries), that remarkable difference observed in the values of the diameter obtained for the (OnlyFemale) shows that in this authorship category, distances among nodes are shortened since 2019. Short diameters correspond to a larger number of leaves, indicating that this network develops an entirely different structure when compared with the other MSTs by authorship category, mainly in 2021 and 2022.

As the female authorship increases in 2022, the MST obtained from the Female category moves from a *star* configuration (in 2019) to a *path* motif (in 2021 and 2022). In so doing, the topological distances between countries are enlarged and the number of poorly connected countries (leaves) decreases. If, conversely, the network of countries has a high percentage of female authorship as in the OnlyFemale MST, the tree would approach a *star* motif as the number of leaves would be enlarged and the corresponding diameter shortened.

Similar important differences are presented in the two trees in Fig.10, when comparing the evolution of the Male authorship, i.e. the articles with at least one male as author, from 2019 to 2022, one sees that:

- a decrease of centrality by China in 2022, likewise in the Female MST

- an increase in the MST diameter in 2022 reveals a more homogeneous degree distribution.

## 4.4   Covid-19 Effects

To identify changes in the gender pattern distribution resulting from the effects of Covid-19 research on IRC, the years 2020 and 2021 are split into two groups: one for for articles including Covid-19 in the WoS "Topic" (i.e. title, abstract keyword or Keywords Plus) and another for the articles not including that topic. The applied query is:

QUERY 02: Articles with EU country authors & 'COVID-19 & similar; years 2020, 2021.

*Source Data Base*: Web of Science, Social Science Citation Index (SSCI).

*Query* : year, article, countries (see QUERY 01 current members of European Union+ England, Wales, North Ireland and Scotland)

Command for years 2020 and 2021: (PY=('year') AND DT=(ARTICLE) AND TS= (COVID-19 OR "Coronavirus" OR "Corona virus"OR "2019-nCoV" OR "SARS-CoV" OR "MERS-CoV" OR "Severe Acute Respiratory Syndrome" OR "Middle East Respiratory Syndrome").

Two different samples were taken for the year of 2021 CovidPapers (N=2,230) and Non-CovidPapers (N=1,112). The categories OnlyFemale and OnlyMale were considered both for Covid and Non-Covid samples. The results show that the OnlyMale MST network induced from CovidPapers (N=554) does not show any reduction in its diameter while the OnlyFemale network (N=348 papers) is disconnected. The exclusive networks OnlyFemale and OnlyMale for 2021 induced from NonCovidPapers were both disconnected, consequently, no MST could be computed. It is likely that the loss of connectivity - being an important issue in the network structure - indicates a heterogeneous distribution of the node's degree.

# 5 Conclusions

There was a fast internationalization of scientific outputs produced by researchers affiliated in EU institutions and simultaneously the slow increase of the share of female participating in this process of IRC. This converge with the results for research in general in EU (European Commission, 2021a) despite the gross enrollment rate of women in higher education which increased from 19% in 2000 to 43% in 2020 among women, but only from 19% to 37% among men (UNESCO, 2022).

Summarizing the answers to the research questions:

1. How have patterns of international research collaboration, both intra-EU and with non-EU countries, developed in the context of EU scientific and innovation policies?

   - In Social Sciences, the share of articles with authors affiliated with EU institutions (includes UK after Brexit) has been increasing, as well as, IRC among EU and non-EU countries. The US is the non-European Union country with the greatest collaboration in research with the European Union countries

   - There was a change in the group of the top 5 countries with the entry of China in 2016 and the withdrawal of Canada in the same year

2. Are there any differences in IRC patterns based on gendered authorship categories?

   - There was a large decrease in the gender gap in 2021

   - Female Exclusive network (OnlyFemale) displays an even greater difference in 2021 and 2022 when compared with either OnlyMale or Total MSTs. Being such a female-exclusive authorship condition the sole determinant of the emergence of that particular shape (far from randomness) in the network structure

3. Has the COVID-19 pandemic crisis altered the extent and nature of gender disparities in IRC, particularly in terms of intensity and collaborative patterns?

   - There was a large decrease in the gender gap in 2021

## 5.1 Concluding remarks

Our study offers several important insights into the gendered dynamics of international research collaboration (IRC). The key concluding remarks are as follows:

- Growth in EU IRC:

  Between 2011 and 2022, there was a substantial increase in IRC in the EU, reflecting the effectiveness of EU policies promoting cross-border scientific cooperation, especially with countries like the USA and China.

- Gender Disparities in IRC Participation:

  Articles with at least one female author were consistently fewer than those with at least one male author. The OnlyFemale category showed structurally different collaboration networks, often with smaller diameter MSTs—indicating more centralized collaboration.

- COVID-19's Dual Impact on Gendered Collaboration:

  The pandemic led to both an increased need for IRC and barriers to it. Female-exclusive networks were more fragmented during COVID-19, suggesting that women researchers faced greater obstacles in maintaining international ties during the crisis.

- Network Analysis via Minimal Spanning Trees (MSTs):

  Using MSTs, the study revealed that networks involving male authors had broader, more equally distributed structures, while those involving only female authors were more star-shaped, indicating centralized collaboration patterns and closeness to randomness.

- Shift in Global Collaboration Patterns:

  The United States maintained its role as the top non-EU research partner, while China significantly rose in prominence, overtaking traditional collaborators like Canada and even England post-Brexit.

## Acknowledgment

- ISEG RESEARCH in Economics and Management, Universidade de Lisboa is financially supported by FCT, I.P., the Portuguese national funding agency for science, research and technology, under the Project UID/06522/2023 FCT (Fundação para a Ciência e a Tecnologia)

- This research was developed under the project GEPINC *Gender Equality Plans for Inclusivity: Engines of Change*, (January 2025-December 2027), European Union's Horizon Europe, Grant Agreement No 101187961.

# Declarations

- Funding

  This work was supported by FCT, I.P., the Portuguese national funding agency for science, research and technology, under the Project UID/06522/2023

- Conflicts of interest/Competing interests

  The authors have no conflicts of interest to declare that are relevant to the content of this article.

- Author Contributions

  E. Fontainha: Conceptualization, Software, Writing - Original draft preparation, Writing-Reviewing and Editing.
  T. Araújo: Methodology, Software, Supervision, Writing-Reviewing and Editing.

- Availability of data and materials Data will be available at a GitHub public repository.

- Code availability

  Code will be available at a GitHub public repository.